# Topological phases in cove-edged and chevron graphene nanoribbons: Geometric structures, $\mathbb{Z}_2$ invariants, and junction states

*Yea-Lee Lee,*[†,‡,¶,§] *Fangzhou Zhao*[†,‡,§] *Ting Cao,*[†,‡] *Jisoon Ihm,*[¶] *and Steven G. Louie*[*,†,‡]

[†]Department of Physics, University of California at Berkeley, Berkeley, California 94720, USA

[‡]Materials Sciences Division, Lawrence Berkeley National Laboratory, 1 Cyclotron Road, Berkeley, California 94720, USA

[¶]Department of Physics, Pohang University of Science and Technology, Pohang, Kyungbuk 37673, Korea

**Corresponding Author**

*S.G.L.: E-mail: sglouie@berkeley.edu

Tel: +1-510-642-1709.

ABSTRACT

Graphene nanoribbons (GNRs) have recently been shown by Cao, Zhao and Louie [Cao, T.; Zhao, F.; Louie, S. G. *Phys. Rev. Lett.* **2017**, *119*, 076401] to possess distinct topological phases in general, characterized by a $\mathbb{Z}_2$ invariant. Cove-edged and chevron GNRs moreover are chemically and structurally diverse, quasi-one-dimensional (1D) nanostructures whose structure and electronic properties can be rationally controlled by bottom-up synthesis from precursor molecules. We derive the value of the topological invariant of the different types of cove-edged and chevron GNRs, and investigate the electronic properties of various junctions formed by these GNRs, as well as such GNRs with the more common armchair or zigzag GNRs. We study the topological junction states at the interface of two topologically distinct segments. For an isolated GNR having two ends of different terminations, topological end states are shown to develop only at the topologically nontrivial end. Our work extends the explicit categorization of topological invariants of GNRs beyond the previously demonstrated armchair GNRs, and provides new design rules for novel GNR junctions as well as future GNR-based nanoelectronic devices.

KEYWORDS graphene nanoribbons, topological phases, heterojunctions, topological states.

The recent discovery[1] of topological phases in graphene nanoribbons (GNRs) together with the theoretical and experimental demonstration of topological quantum engineering[2,3] of GNR systems have paved a new path towards rational design and control of the electronic properties of bottom-up GNRs using their topological properties. Moreover, the past decade has seen the rise of bottom-up synthesis of a diversity of GNRs[4-12] providing a wealth of systems for the further studies and possible applications.

The bottom-up GNR structures are derived from self-assembly of small molecular building blocks, and therefore have atomically-smooth, well-defined edges and terminations, posing a significant advantage over traditional top-down GNRs.[13-17] The versatility of the molecular building blocks enables atomically precise designs of diverse types of GNRs, including the manipulation of the width and edge structures,[4,10,11] and the atomic doping and chemical functionalization in the interior[18] or on the edges[8,19,20] of the GNRs. As such, having a large variety of GNR structures realized by bottom-up synthesis has led to tremendous promise in nanoelectronics and other technologies,[6] especially in band gap engineering,[5,20] one-dimensional (1D) heterojunctions,[8,9,20] and so on. Since topology-induced localized states exist at junctions between GNRs of different topological phases (owing to the bulk-boundary correspondence), a superlattice of such junctions results in new topological derived bands[1], leading to the concept of topological quantum band engineering.[2,3]

We note that the topological phases of 1D crystals have additional rich physical behaviors from those of two-dimensional (2D) or three-dimensional (3D) systems. For example, in 2D quantum spin Hall insulators,[22,23] the topological phase is protected by time-reversal symmetry and the topological invariant characterizing it corresponds to integration of the Berry curvature of the occupied energy bands for each spin species in the Brillouin zone (BZ).[24-27] The topological invariant does not depend on the choice of the unit cell. However, in 1D GNR systems, in additional to time reversal symmetry, the topological phase depends on the presence of spatial symmetries such as inversion or mirror symmetry.[1,28] In these systems, the topological invariant is characterized by the sum of the intercell Zak phase of the occupied bands, which is an integral of the Berry connection in the 1D BZ and is quantized to 0 or $\pi$ (mod $2\pi$)[1,27-29] for a unit cell with certain spatial symmetry. Here the intercell Zak phase does not depend on the choice of spatial origin, but does depend on the choice of the unit cell that is physically dictated by a structure's termination geometry. Hence for 1D crystals, as in the well-known case of polyacetylene,[30] the electronic topology has an extra and rich dependence on the system's geometric boundary which can be used to manipulate the topological properties of the material.[31]

Previous study[1] has derived explicit expressions for the value of the $\mathbb{Z}_2$ invariant of armchair graphene nanoribbons (AGNRs) that depend on both the ribbon width and the unit cell that is chosen to be commensurate to the geometric boundary of the system of interest. A unit cell of an AGNR that supports inversion and/or mirror symmetry can have zigzag, zigzag', or bearded carbon-bond configuration at the unit cell boundary, depending on the AGNR's terminating geometry. For example, the end termination of an N = 7 AGNR (where N is the number of rows of carbon atoms forming the ribbon width) can be either zigzag or zigzag', giving rise to a unit cell that has either a $\mathbb{Z}_2$ invariant of 1 or 0, respectively[1].

In this Letter, we extend the study of GNR topological phases[1] to cove-edged[32] and chevron[4] GNRs, which have been recently synthesized, and we determine the dependence of their topological phases on width, edge, and boundary structures. Density functional theory (DFT) calculations as implemented in the Quantum Espresso package,[33] Wannier function (WF) method,[34-36] and tight-binding models[1,27,37] are used to calculate the $\mathbb{Z}_2$ invariants of these highly interesting and more complex GNRs. We then explore the consequences of having topological phase differences at heterojunctions or ends of finite segments of such GNRs.

The structures of cove-edged GNRs are essentially zigzag GNRs with periodic carbon atom vacancies on both edges.[32] (See Fig. 1.) Depending on the vacancy positions and density, the cove-edged GNRs can be semiconducting or nearly metallic (with a tiny gap) within the local spin-density functional approximation (DFT-LSDA). We consider here a prototypical family of cove-edged GNRs where one carbon vacancy and one hexagonal carbon ring alternatively occur along each edge. Among all possible cove-edged GNRs, this family gives the smallest unit cell size along the 1D periodic direction and has been recently synthesized by the bottom-up process.[32] This family of short-period cove-edged GNRs can be further categorized into three types – determined by its width specified by N (the number of zigzag chains forming the width of the parent zigzag ribbon) and the relative positions of the vacancies placed across the opposite edges (Fig. 1).[38] For N = odd, there is only one type of structure.[32] (See Fig. 1(a)) In contrast, when N = even, there are two distinct types of structures: the carbon vacancies on the opposite edges can either be directly facing each other (named N = symmetric-even) or be staggered (named N = asymmetric-even) as depicted in Fig. 1 (b) and (c), respectively.

The fully relaxed atomic structures of free-standing cove-edged GNRs show that the hexagonal carbon rings on the edges (Fig. 1(a)) are either tilted upward (pink) or downward (blue) from the GNR plane. This is because the edge carbon atoms are passivated by hydrogen atoms, as in the experimental bottom-up samples, and the steric hindrance between neighboring hydrogen atoms leads to an out-of-plane bending of the edge carbon rings. For the N = 5 cove-edged GNR in Fig. 1(a), the edge carbon atoms have a maximum height of ±1.2Å from the center plane in our DFT calculations. We did not find any spin polarization at the DFT-LSDA level for the cove-edged GNRs studied, even though the parent zigzag GNRs have spin-polarized edge states.[39]

The short-period cove-edged GNRs of types N = odd and N = symmetric-even have sizable DFT-LSDA energy gaps that increase as the GNR gets narrower. By contrast, the cove-edged GNRs of type N = asymmetric-even show almost zero band gap regardless of its width. Comparing between the two types of N = even cove-edged GNRs, shifting of the vacancy positions on one edge relative to those on the other edge gives rise to dramatically different electronic structures. As shown in the DFT-LSDA band structure of the N = 6-asymmetric cove-edged GNR (lower panel of Fig. 1(c)), the linear band dispersion near the Fermi energy that originates from the Dirac-like band dispersion at the K point of graphene folds to the zone center of the 1D BZ. This feature of the band structure for N = asymmetric-even survives in our LSDA and fully relativistic calculations including spin-orbit interaction.[40]

Knowing the electronic structure, we can evaluate the Zak phase[28] for the cove-edged GNR with different shapes and unit cells. The Zak phase of the $n$th band $\gamma_n$ is the integral over the BZ of the Berry connection written as

$$\gamma_n = i \int_{BZ} d\bm{k} \, \langle u_{n\bm{k}} | \nabla_{\bm{k}} u_{n\bm{k}} \rangle, \qquad (1)$$

where $|u_{n\bm{k}}\rangle$ is the periodic part of the Bloch function of the $n$th band at momentum $\bm{k}$, and the total Zak phase for a given 1D insulator is the sum of $\gamma_n$ over the occupied band complex. Once we obtain $|u_{n\bm{k}}\rangle$ (e.g., from DFT calculations or tight-binding model Hamiltonians), we can use various evaluation schemes for the Zak phase such as the WF method,[28,34-46] the Fu-Kane method,[27] or numerically integrating the Berry connection over the 1D BZ.[37]

In the WF method, we calculate the Wannier centers (WCs) which are the expectation values of the position operator $\hat{\bm{r}}$ for the WFs. If inversion or mirror symmetry exists for a given unit cell, the vector sum of the WCs for all the occupied bands (modulo translational vectors) is either located exactly at the center or at the boundaries of the unit cell.[28] The corresponding total Zak phase is 0 or π, and the topological invariant $\mathbb{Z}_2$ (defined as $(-1)^{\mathbb{Z}_2} = e^{i\sum_n \gamma_n}$)[1] is 0 or 1, respectively.

For example, we can arrive at the value of the topological invariant of the N = 5 cove-edged GNR system from the centers of the WFs shown in Fig. 1(e). The red dots represent the positions of the π-electron WCs. The sum of the positions of all the occupied π-WCs is located at the center of the unit cell as shown in Fig. 1(e), which indicates that the total Zak phase is 0 and $\mathbb{Z}_2$ = 0.[28] (See Supporting Information (SI).) If we shift the defining boundary of the unit cell by ¼ of a primitive translation vector, inversion symmetry still holds, but the sum of the positions of the π-electron WCs is now located at the boundary of the new unit cell. As a result, the topological invariant $\mathbb{Z}_2$ becomes 1 for the new, shifted unit cell.

We also evaluate the value of the topological invariant of GNRs from the electronic states directly with two other methods. The electronic states can be obtained either via a tight-binding method for the $p_z$ orbitals of the carbon atoms[41,42] or via the DFT approach. Once we have the electronic structure, since our 1D system has inversion and/or mirror symmetry, the Fu-Kane method[27] may be used to determine the value of $\mathbb{Z}_2$ by examining the parity of the occupied wave functions at parity-invariant k points. Alternatively, the integrand of the Eq. (1) may be numerically evaluated at discrete k points in the 1st BZ for a direct computation of the integral.[37] These calculational details are discussed in SI.

The values of the $\mathbb{Z}_2$ invariant for various short-period cove-edged GNRs are presented in Table 1, which depend on their widths and the choice of unit cells (which as discussed above is determined by the terminal structure). We use the nomenclatures established above to present and categorize the topological phases of these cove-edged GNRs. For cove-edged GNR with N = odd and N = symmetric-even, we select six representative unit cells (commensurate to different structural terminations of the ribbon) that have inversion and/or mirror symmetry. For N = odd cove-edged GNRs, unit cells with armchair and armchair' cell boundary shapes have trivial and nontrivial topological phases, respectively, regardless of the GNR widths. In comparison, for the other four types of cell boundary shapes (zigzag, zigzag', bearded, and bearded') considered in this work, the $\mathbb{Z}_2$ invariant for the N = odd cover-edged GNRs changes cyclically with a periodic of 8 on N(see Table 1). For example, for N = $8p + 5$ ($p$ is an integer), such a cove-edged GNR can have all the six cell boundary shapes. The armchair and armchair' shapes lead to $\mathbb{Z}_2 = 0$ and 1, respectively. For the other unit cell boundary shapes, a zigzag (zigzag') 60°-tilted unit cell has $\mathbb{Z}_2 = 1$ (0), and a bearded (bearded') 120°-tilted unit cell has $\mathbb{Z}_2 = 1$ (0). The $\mathbb{Z}_2$ for all other N = odd cove-edged GNRs are similarly presented in the Table. For N = symmetric-even, spatial symmetry only exists in zigzag- (zigzag'-) shaped unit cells for N = $8p + 2$ and $8p + 6$, and in bearded- (bearded'-) shaped unit cells for N = $8p$ and $8p + 4$. The zigzag- (zigzag'-) shaped boundaries result in $\mathbb{Z}_2 = 0$ (1) for N = $8p + 2$, and $\mathbb{Z}_2 = 1$ (0) for N = $8p + 6$. The bearded- (bearded'-) shaped boundaries have $\mathbb{Z}_2 = 1$ (0) for N = $8p$, and $\mathbb{Z}_2 = 0$ (1) for N = $8p + 4$. For N = asymmetric-even, the cove-edged GNRs are virtually gapless within DFT-LSDA; so, we do not consider their topological invariants in this study.

Next, we investigate the electronic structure and $\mathbb{Z}_2$ of the chevron GNRs.[4,21,43] The geometric structures of chevron GNRs are related to those of armchair GNRs. Three specific kinds of chevron GNRs with different unit cell shapes are considered in this work: regular chevron GNR,[4] extended-chevron GNR,[43] and binaphthyl-chevron GNR[21] (Fig. 2). They all have the narrowest width portions in the unit cell formed from N = 6 armchair GNR. These chevron GNRs have been recently synthesized through bottom-up self-assembly of molecular building blocks.[4,21,43] The extended-chevron GNR and binaphthyl-chevron GNR are laterally extended structures that can be viewed as a regular chevron GNR with additional hexagonal carbon rings and additional two rows of carbon in the elbow positions, respectively. (See Fig. 2.) The band structures of the three types of chevron GNRs are shown in Fig. 2. Their band gaps, evaluated by DFT-LDA, are 1.18 (binaphthyl), 1.37 (extended), and 1.59 (regular) eV. The band gaps of extended- and binaphthyl- chevron GNRs are smaller than their parent regular chevron GNR because increasing effectively the ribbon widths by the additional carbon atoms leads to weaker quantum confinement.[21,43]

We have calculated the $\mathbb{Z}_2$ invariants of these chevron GNRs for the two different unit cell shapes indicated in Fig. 2. For regular chevron and binaphthyl-chevron GNRs, $\mathbb{Z}_2 = 0$ for the 60° or 120° unit cells with armchair shape at the unit cell boundary, and $\mathbb{Z}_2 = 1$ for the 90° unit cells with zigzag shape at the unit cell boundary. But extended-chevron GNRs with 60°, 90° and 120° unit cells all belong to the topological trivial class, i.e., $\mathbb{Z}_2 = 0$.

We next explore various GNR heterojunctions or finite segments that show junction states or end states, resulting from the topological nature of the electronic structure of their components. Guided by the values of $\mathbb{Z}_2$ of GNRs with various widths and terminations, we construct a number of different GNR heterojunctions – consisting of combination of an N = 7 AGNR, an N = 5 cove-edged GNR, or a regular chevron GNR, at which two topologically equivalent or inequivalent segments are connected together. The emergence of topologically-induced junction states for the latter class of junctions has been confirmed for all these structures.

Fig. 3(a) shows two heterojunctions made from an N = 7 AGNR and an N = 5 cove-edged GNR with tilting angle of 30°. The two junctions are structurally different, corresponding to either 1) a zigzag ($\mathbb{Z}_2 = 1$) termination or 2) a zigzag' ($\mathbb{Z}_2 = 0$) termination of an N = 7 AGNR joining with a zigzag ($\mathbb{Z}_2 = 1$) termination of an N = 5 cove-edged GNR. For case 2), as expected, a topologically-induced junction state emerges and is localized at the interface between the two topologically inequivalent segments (the zigzag'-terminated N = 7 AGNR and zigzag-terminated N = 5 cove-edged GNR junction). In comparison, there is no in-gap state in case 1), the heterojunction formed between the zigzag-terminated N = 7 AGNR and zigzag-terminated N = 5 cove-edged GNR. Fig. 3(b) depicts that an N = 7 AGNR/regular chevron GNR heterojunction can also be formed in two different configurations. Again, by bulk-boundary correspondence, a topological junction state arises at the zigzag'-terminated N = 7 AGNR/zigzag-terminated chevron GNR junction; but, in the other case of zigzag-terminated N = 7 AGNR/zigzag-terminated chevron GNR junction, there is no in-gap state. The square of wave function of the junction states from DFT-LDA, integrated over the direction perpendicular to the ribbon plane [in units of $1/(a.u.)^2$], are shown in the bottom panel of Fig. 3(a) and 3(b). Fig. 3(c) and (d) show the DFT-LDA density of states (DOS) of the two systems, where a broadening of 0.03 eV is used. (The structures for which the DOS are evaluated are shown in Fig. S1 and S2 in the SI. We employed a finite structure with a length of 7.94 nm for the N =7 AGNR/N=5 cove-edged GNR heterojunction and a superlattice structure with a lattice parameter of 10.23 nm for the N = 7 AGNR/chevron GNR heterojunction.) Because we have a finite length for each of the GNR segments in the calculation, the DOS from the bulk regions away from the junction does not dominate so that the junction states are visible in the DOS plot.

Our calculations show that the bulk bands of the N = 7 AGNR/N = 5 cove-edged GNR heterojunctions (for both cases in Fig. 3(a)) are aligned in the form of a type-I heterojunction (see SI). The energy band gaps within DFT-LDA of the individual N = 7 AGNR and the N = 5 cove-edged GNR are 1.6 eV and 0.93 eV, respectively. The type-I band alignment gives the wave functions of the low-energy conduction and valence band states away from the band gap of the combined systems (not counting the in-gap state) residing in the cove-edged GNR region, and the ones of the higher-energy states are extended throughout the whole system. On the other hand, a regular chevron GNR has a DFT-LDA band gap of 1.59 eV which is almost the same as that of the N = 7 AGNR. Our results show that the bulk band states of the N = 7 AGNR/regular chevron GNR heterojunction systems (for both cases in Fig. 3(b)) are all extended throughout the system, i.e., it is a type-I alignment between two semiconductors with nearly identical band gaps.

As mentioned before, the value of the $\mathbb{Z}_2$ invariant of a 1D system is related to how the end of the system terminates. Thus, we can design a finite GNR segment with different $\mathbb{Z}_2$ at the two ends. We now consider the topologically-induced end states of individual finite segments of N = 5 cove-edged GNRs and of regular chevron GNRs, for which one end has a terminating unit cell of $\mathbb{Z}_2 = 0$ and the other end has a terminating unit cell of $\mathbb{Z}_2 = 1$ (Fig. 4(a) and (b)). From bulk-boundary correspondence, an end state would arise at the end with $\mathbb{Z}_2 = 1$ since vacuum is topologically trivial, and there would not be an end state at the other end with $\mathbb{Z}_2 = 0$. Of course, at the end of a ribbon, it is also possible that localized states occur not from a difference in topological phases, but from a modification of the potential seen by the electron, such as broken chemical bonds, imperfection/distortion of the carbon rings, etc. Even so, if the commensurate unit cell to a specific end structure is inversion/mirror symmetric, the total number of end states at the GNR-vacuum interface should follow the $\mathbb{Z}_2$ invariant rule – an even or odd number of localized states at the end with $\mathbb{Z}_2 = 0$ or $\mathbb{Z}_2 = 1$, respectively. We illustrate in Fig. 4 the character of the topological end states for several such finite segments with plots of their wavefunctions from explicit DFT-LDA calculations.

As discussed in Ref. 1, the topologically-induced junction and end states are robust against perturbations such as small displacements in atomic positions and/or changes in the atomic potentials (e.g., within tight-binding formalism, changes in the hopping integrals and onsite energies); these have been explicitly confirmed in our calculations for the systems studied. Since only the π electrons are relevant to the topological phase of the GNRs, perturbations to the edge structure that retain the integrity of the band structure of the π-electron system should not change the junction or end states. However, since the topological invariants are determined by the spatial symmetry of the unit cell that is commensurate to the termination structure, a change in termination geometry for the π electrons will change a segment's topological phase or making it ill-defined.

Due to electron-electron interaction, localized topological junction/end states in the band gap in general would have a Coulomb energy penalty for occupancy of a second electron.[1] For example, if the onsite repulsion is large, in a spin-polarized calculation, the energy level corresponding to the topological-induced junction states may be split into spin-up and spin-down polarized states, with one located below the Fermi level and the other above the Fermi level.

In summary, the electronic structure and topological phase of various cove-edged and chevron GNRs are investigated, which expand the categorization of the topological phases of GNR systems going beyond the common armchair and zigzag GNRs[1]. Depending on the ribbon width, edge shape, and end terminating geometry, we show that distinct topological phases exist in these more complex GNR systems, leading to a variety of possibilities for designing heterojunctions and finite segments that exhibit topologically-induced junction/end states. These topological junction and end states may be developed for various applications, including e.g., quantum topological band engineering and quantum information.

**Methods.** We have performed first-principles DFT calculations within the local (spin) density approximation (L(S)DA) for various GNR species, employing supercell slab geometries as implemented in the Quantum Espresso Package.[33] Norm conserving pseudopotentials with a planewave energy cut-off of 60 Ry are used, and a Monkhorst $k$-mesh is chosen as 18×1×1 for the cove-edged GNRs, 6×1×1 for the chevron GNRs, 4×1×1 for the heterojunction of $N = 7$ AGNR/chevron GNR in a superlattice structure, and 1×1×1 for the finite length heterojunctions and finite individual segments studied. All structures were fully relaxed until the force on each atom is smaller than 0.005 eV/Å in the bulk of the GNRs, and smaller than 0.026 eV/Å for atoms at the boundaries of heterojunctions and finite segments. Maximally localized Wannier functions for the cove-edged GNR of $N = 5$ were obtained employing the Wannier90 package.[44]

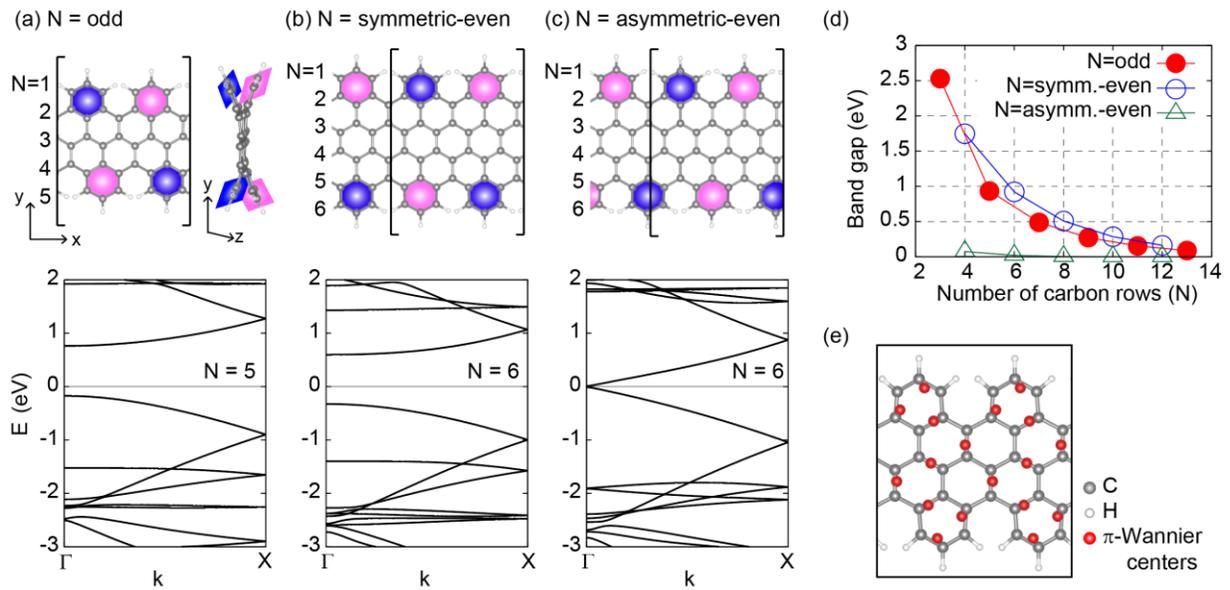

**Figure 1.** Schematic structures (top panels) and band structures calculated using DFT-LSDA (bottom panels) of short period cove-edged GNRs of (a) N = 5 (odd), (b) N = 6 (even) with symmetric edges (aligned facing vacancies), and (c) N = 6 (even) with asymmetric edges (misaligned facing vacancies) across the ribbon. The cove-edged GNRs are periodic along the x direction, and the side view in (a) illustrates the structural deformation along the direction perpendicular to the GNR plane. The unit cell of each GNR studied is shown by the black bracket. The gray and silver balls denote carbon and hydrogen atoms, respectively. The pink and blue areas denote edge carbon rings tilting upward and downward from the ribbon plane, respectively. (d) DFT-LSDA band gap of the three types of cove-edged GNRs in (a)-(c). For N =

asymmetric-even cove-edged GNRs, the DFT-LSDA band gap is close to 0. (e) Wannier centers of the π-electron Wannier functions of an N = 5 cove-edged GNR are denoted by red dots in the unit cell.

**Table 1.** Calculated values of $\mathbb{Z}_2$ invariants of short-period cove-edged GNRs for Different Widths N (mod 8) and Shape of the Unit Cells. Unit cell shapes studied are armchair, armchair', zigzag, zigzag', bearded, and bearded'. The atomic structure in the second row of the table is illustrated with an N = 5 cove-edged GNR, with gray and silver balls denoting carbon and hydrogen atoms, respectively. For N = odd, the unit cell boundary with a zigzag(') and bearded(') shapes forms a 60° and 120° angle with respect to the ribbon axis, respectively. For N = even, only the symmetric carbon vacancy structures which have sizable DFT-LSDA band gap are considered here. For unit cells with atomic structure that does not have spatial symmetry, the Zak phase is not quantized, so that a meaningful $\mathbb{Z}_2$ invariant does not exist and this is denoted by "not applicable" (N/A).

| Termination type<br>N=width<br>($p$ integer) | | Armchair 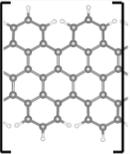 | Armchair' 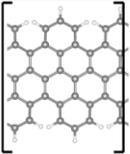 | Zigzag 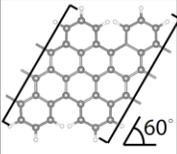 | Zigzag' 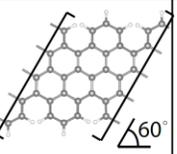 | Bearded 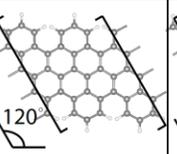 | Bearded' 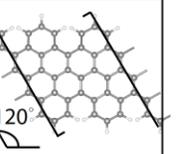 |
|---|---|---|---|---|---|---|---|
| odd | $8p + 1$ | 0 | 1 | 0 | 1 | 0 | 1 |
| | $8p + 3$ | 0 | 1 | 0 | 1 | 1 | 0 |
| | $8p + 5$ | 0 | 1 | 1 | 0 | 1 | 0 |
| | $8p + 7$ | 0 | 1 | 1 | 0 | 0 | 1 |
| even | $8p + 2$ | N/A | | 0 | 1 | N/A | |
| | $8p + 6$ | | | 1 | 0 | | |
| | $8p$ | | | N/A | | 1 | 0 |
| | $8p + 4$ | | | | | 0 | 1 |

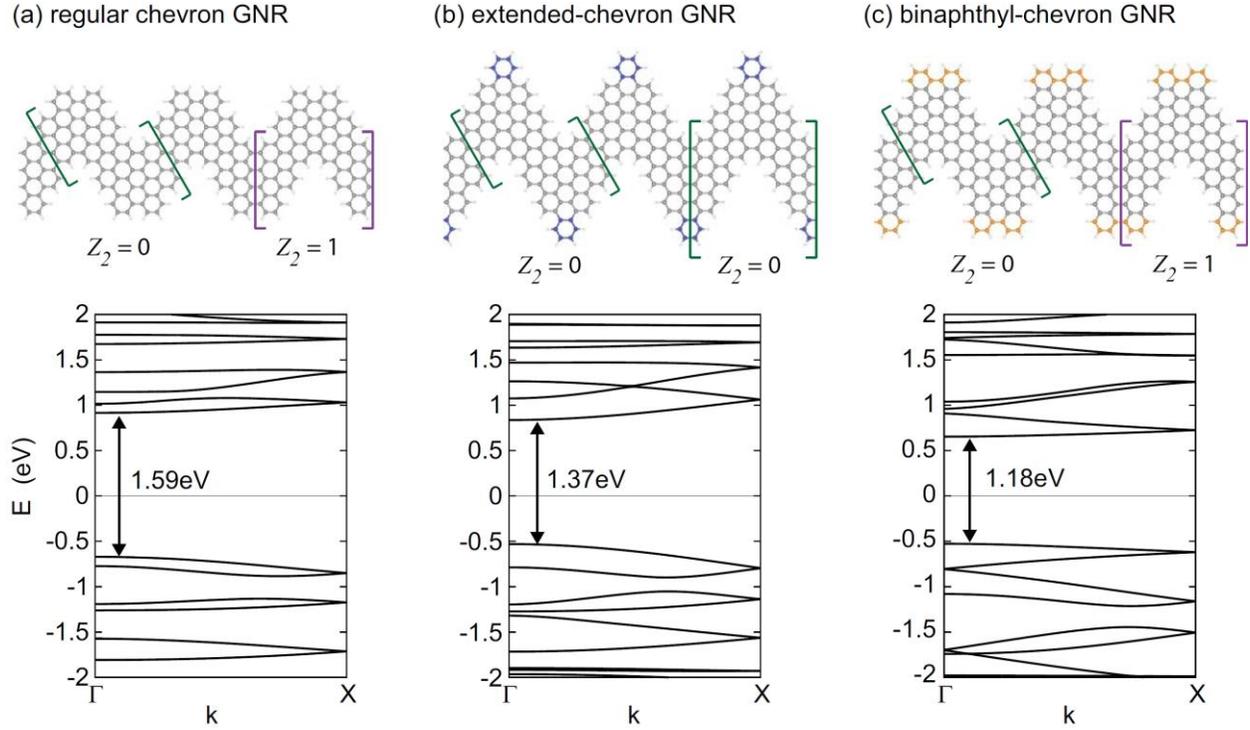

**Figure 2.** Schematic structures (top) and DFT-LDA band structures (bottom) of (a) regular chevron, (b) extended-chevron, and (c) binaphthyl-chevron GNRs. Gray and silver balls denote carbon and hydrogen atoms, respectively. The structures of extended- and binaphthyl-chevron GNRs are laterally extensions of a regular chevron GNR with additional hexagonal carbon rings (blue) and additional two rows of carbon (orange) in the elbow positions, respectively. They all have the narrowest width portions in the unit cell formed from N = 6 armchair GNR. The green and purple brackets show the unit cell shapes corresponding to topologically trivial and nontrivial phases ($\mathbb{Z}_2$ = 0 and 1), respectively. The DFT-LDA band structures show a semiconducting gap of 1.59, 1.37, and 1.18 eV in regular chevron, extended-chevron and binaphthyl-chevron GNRs, respectively.

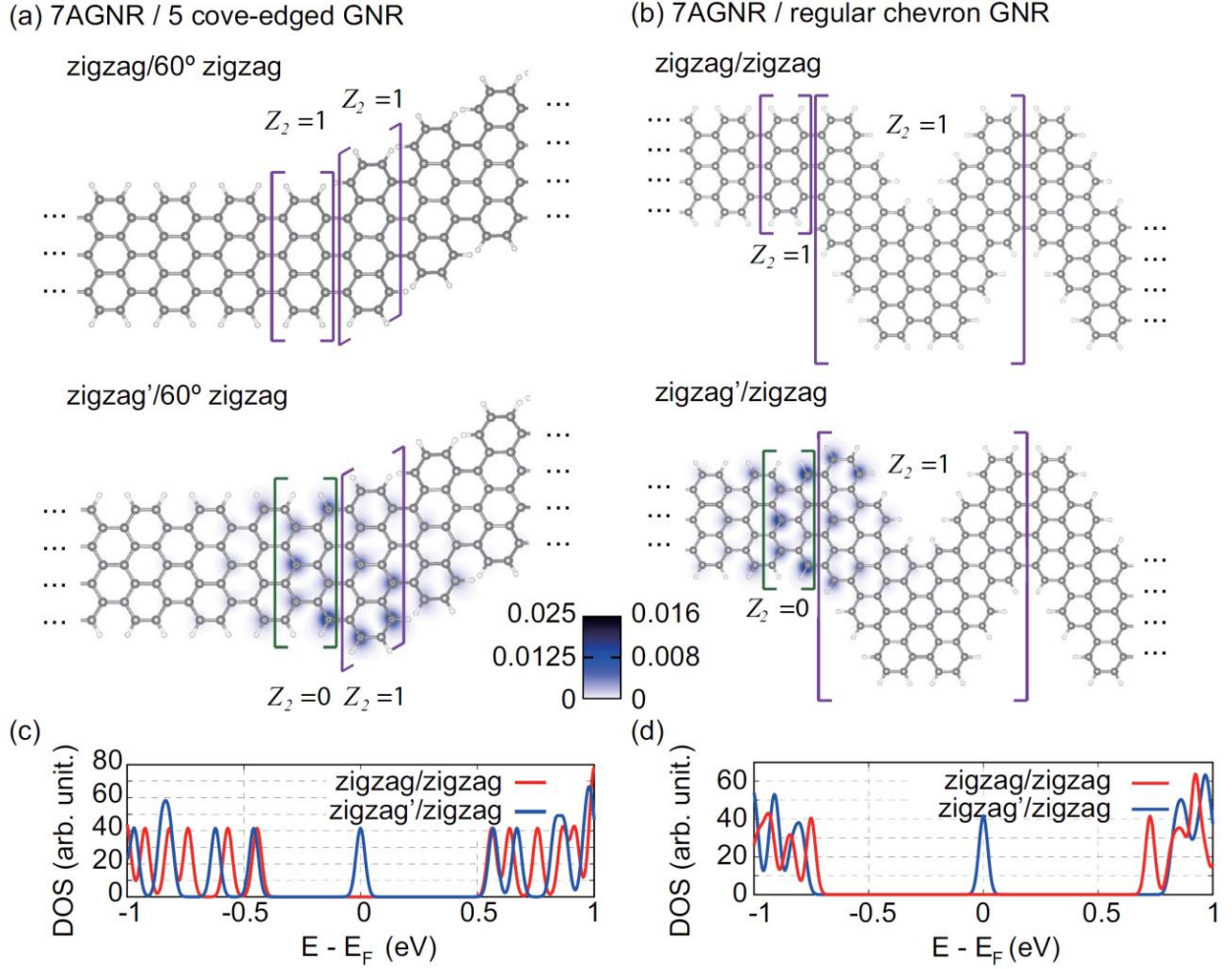

**Figure 3.** GNR heterojunctions formed with (a) an N = 7 AGNR and an N= 5 cove-edged GNR (finite length of 7.94 nm), and (b) an N = 7 AGNR and a regular chevron GNR (superlattice with lattice constant a = 10.23 nm). The upper panel and lower panel in (a) and (b) show heterojunctions between two topologically equivalent segments (zigzag/zigzag interface) and inequivalent segments (zigzag'/zigzag interface), respectively. These are zoom-in schematic structures near the junctions. The full structures used in the calculations are shown in SI. The gray and silver balls represent the carbon and hydrogen atoms, and the green (purple) brackets show the unit cell with $\mathbb{Z}_2 = 0$ (1). The color scale shows the square of wave function of the topological junction states integrated over the direction perpendicular to the ribbon plane [in units of $1/(a.u.)^2$]. The density of states of (c) the N = 7 AGNR/N = 5 cove-edged GNR heterojunctions and (d) the N = 7 AGNR/chevron GNR heterojunctions, with zigzag/zigzag (red curves) and zigzag'/zigzag (blue curves) interfaces, where a broadening of 0.03 eV is used. Topologically-induced junction states emerge at the zigzag'/zigzag interfaces.

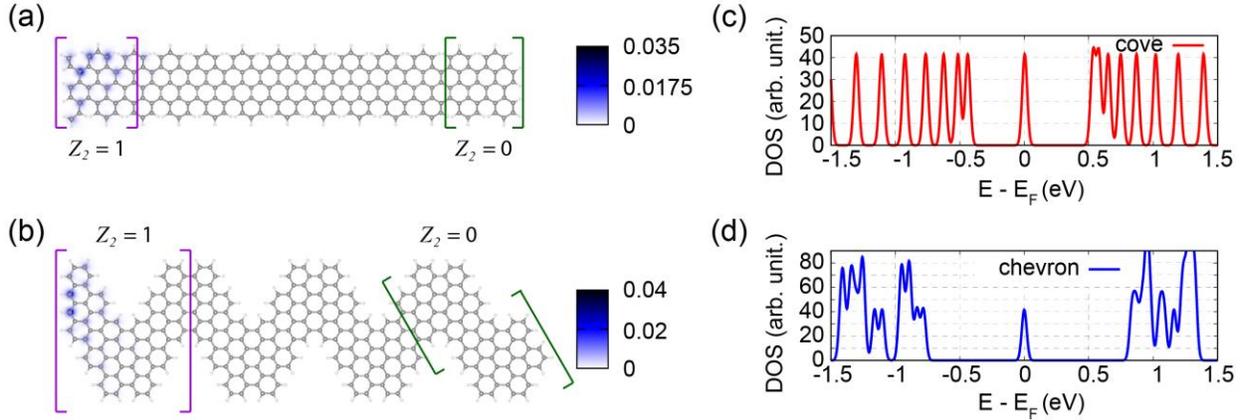

**Figure 4.** Finite GNR segments with different end terminations for (a) an N = 5 cove-edged GNR (a finite length of 6.21 nm) and (b) a regular chevron GNR (a finite length of 6.67 nm). Both isolated GNR segments consist of one end with $\mathbb{Z}_2 = 1$ and the other end with $\mathbb{Z}_2 = 0$. The gray and silver balls represent the carbon and hydrogen atoms, respectively, and the green (purple) bracket shows the unit cell with $\mathbb{Z}_2 = 0$ (1) at a particular end. The color scale shows the charge density of the topologically-induced end state at the end with $\mathbb{Z}_2 = 1$. The charge density of the end state is integrated over the out-of-plane direction [in units of $(1/a.u.)^2$]. The density of states from DFT-LDA calculation of (c) cove-edged GNR segment and (d) chevron GNR segment with structures shown in (a) and (b) respectively are plotted, where an energy broadening of 0.03 eV is used. A two-fold degenerate topologically-induced end level appears at mid-gap (or at the Fermi level since it is half occupied) for both GNR segments.

## ASSOCIATED CONTENT

**Supporting Information**

The following file is available free of charge.

Supporting Information on topological phases in cove-edged and chevron graphene nanoribbons (PDF)

## AUTHOR INFORMATION

**Corresponding Author**

*sglouie@berkeley.edu

**Author Contributions**

§Y.-L.L. and F.Z. contributed equally to this work.

**Notes**

The authors declare no competing financial interest.

## ACKNOWLEDGMENT

This study was supported by NSF Grant No. DMR1508412, the NSF Center for Energy Efficient Electronics Science (E3S, NSF Grant No. ECCS-0939514), and the Office of Naval Research MURI under Award No. N00014-16-1-2921. Computational resources were provided by the DOE at Lawrence Berkeley National Laboratory's NERSC facility and the NSF through XSEDE resources at NICS.

## ABBREVIATIONS

GNR, graphene nanoribbon; SPT, symmetry-protected topological; DFT, density functional theory; WF, Wannier function; WC, Wannier center.